\documentclass{article}

\begin{document}

\rightline{July 2006}

\vskip 2cm

\centerline{\bf \huge Generalised leptonic colour}

\vskip 2.2cm

\centerline{\large R. Foot and R. R. Volkas\footnote{ E-mail address:
rfoot@unimelb.edu.au, r.volkas@physics.unimelb.edu.au}}

\vspace{0.7cm}

\centerline{\large \it School of Physics,}

\vspace{1mm}

\centerline{\large \it Research Centre for High Energy Physics}

\vspace{1mm}

\centerline{\large \it University of Melbourne,}

\vspace{1mm}

\centerline{\large \it Victoria 3010 Australia}

\vspace{1.2cm}

It is conceivable that there is an $SU(N)_{\ell}$ `colour' gauge group for
leptons, analogous to the gauged $SU(3)_q$ colour group of the quarks. 
The standard model emerges as the low energy effective theory when
the leptonic colour is spontaneously broken. The simplest such generalised leptonic 
colour models are constructed.
We show that the see-saw mechanism for small neutrino masses, along
with the theoretical constraint of electric charge quantisation, 
suggests that the models 
with $N = 3, 5, 7$ are the theoretically most promising cases.
A striking feature of generalised leptonic colour is the physics 
associated with the extra leptonic degrees of freedom -- the liptons. These
particles can potentially be discovered at future colliders, such as the LHC, 
making the idea testable in the near future.

\newpage

Under the $SU(3)_q \otimes SU(2)_L \otimes U(1)_Y$ gauge group, 
the quantum numbers for one generation of standard 
model fermions are
\begin{eqnarray}
f_L \sim (1, 2, -1),\quad e_R \sim (1, 1, -2), \quad \nu_R \sim (1, 1, 0),
\nonumber \\
Q_L \sim (3, 2, 1/3),\quad d_R \sim (3, 1, -2/3), \quad u_R \sim (3, 1, 4/3),
\label{1}
\end{eqnarray}
where we have included a $\nu_R$ to enable the neutrinos to gain mass via the
see-saw mechanism.

The $U(1)$ quantum numbers are not a complete theoretical mystery, as 
they can be derived 
from gauge invariance at the quantum level (anomaly cancellation) and
classical level (gauge invariance of the Yukawa Lagrangian) \cite{cq}. 
However, there is no
overall theoretical explanation for the required pattern, or for the 
particular gauge group chosen.

Within the pattern, Eq.(\ref{1}), there is a certain similarity 
between quarks and leptons. Of course, it is not known whether this
similarity should be viewed as some type of coincidence or has some deeper 
significance.  An obvious possibility is that the similarity is the result of a
spontaneously broken exact symmetry, as in, for example, 
the $SU(4) \otimes SU(2)_L \otimes SU(2)_R$ Pati-Salam model \cite{pati}, 
or in the $SU(3)_{\ell} \otimes SU(3)_q \otimes SU(2)_L \otimes U(1)_X$
quark-lepton symmetric model \cite{ql}.
The former model features a 
continuous gauge symmetry relating the quarks and the leptons, 
while the latter model has a discrete $Z_2$ symmetry 
interchanging the quarks with
the leptons; in both cases the symmetry is spontaneously 
broken.\footnote{Both of these models can be grand
unified:  Pati-Salam can be extended to $SO(10)$ \cite{georgi}, while the quark-lepton
symmetric model can be extended to quartification $SU(3)^4$ \cite{quart}.} 
Alternatively, the similarity of the quarks and the leptons 
might simply imply a closer correspondence between the quarks 
and leptons, but not an exact Lagrangian symmetry.

In this vein, we shall generalise the quark-lepton symmetric model \cite{ql} 
by exploring the possibility that there is a leptonic colour analogue of 
the $SU(3)_q$, but we do not demand that there be an exact Lagrangian symmetry. 
Specifically, we explore the hypothesis that 
the leptons transform under an $N$-plet of leptonic $SU(N)_{\ell}$ 
colour, so that the gauge group of the model is\footnote{
A model with this gauge group has been earlier discussed in Ref.\cite{flvtech}. 
However, that was a technicolour model where the extra leptonic 
degrees of freedom
were massless and confined by an unbroken $SU(N-1)$ subgroup
of $SU(N)_{\ell}$ at a high scale -- dynamically breaking electroweak 
gauge symmetry. 
In view of the well-known difficulties of technicolour
with the electroweak precision data we here assume standard gauge symmetry
breaking via Higgs scalars.}
\begin{eqnarray}
SU(N)_{\ell} \otimes SU(3)_q \otimes SU(2)_L \otimes U(1)_X\ .
\label{gauge}
\end{eqnarray}
Gauge anomaly cancellation implies the generation pattern
\begin{eqnarray}
F_L &\sim & \left(N, 1, 2, {-3y\over N}\right), \quad  
E_R \sim \left(N, 1, 1, {-3y \over N} -1\right), \quad 
N_R \sim \left(N, 1, 1, {-3y\over N} +1\right), \nonumber \\
Q_L & \sim & (1, 3, 2, y), \quad d_R \sim (1, 3, 1, y-1), \quad u_R \sim (1, 3, 1, y+1),
\label{2}
\end{eqnarray}
where we have also enforced classical constraints from the standard Yukawa Lagrangian:
\begin{eqnarray}
{\cal L}_{yuk} = \lambda_1 \bar F_L \phi E_R + \lambda_2 \bar F_L \phi^c N_R
+ \lambda_3 \bar Q_L \phi d_R + \lambda_4 \bar Q_L \phi^c u_R \ + \ H.c.
\end{eqnarray}
In this Lagrangian, $\phi \sim (1, 1, 2, +1)$ is the standard model 
Higgs doublet responsible for electroweak symmetry breaking and standard fermion 
masses. [Note that we have normalised the $U(1)_X$ charge of $\phi$ to 1 without losing 
generality.]

The parameter $y$ in Eq.(\ref{2}) is undetermined from the theoretical 
constraints of anomaly cancellation and classical invariance of the Yukawa Lagrangian. It
exists because there are two independent Abelian charges allowed by the classical
and quantum gauge invariance constraints, so if one chooses to gauge only
one such charge it can be an arbitrary linear combination of the two.  A similar
phenomenon arises in the standard model with Dirac neutrinos, where both standard
hypercharge and $B-L$ are classically and quantum mechanically allowed \cite{cq}.

The exotic leptonic degrees of freedom can be given mass from the scalar, $\chi$,
via the Yukawa Lagrangian
\begin{eqnarray}
{\cal L}_{yuk} = \lambda \bar F_L \chi (F_L)^c + \lambda' \bar E_R \chi
(N_R)^c  \ + \ H.c.
\label{yuk}
\end{eqnarray}
In order for all of the exotic leptonic degrees of freedom to gain mass, 
$\chi$ needs to be in the antisymmetric $N(N-1)/2$ dimensional representation of
$SU(N)$. Gauge invariance of the above Lagrangian then implies that $\chi$ has the
gauge quantum numbers
\begin{eqnarray}
\chi \sim \left(\overline{{N(N-1) \over 2}}, 1, 1, {-6y \over N}\right).
\end{eqnarray}
The Higgs potential terms just involving the $\chi$ are
\begin{eqnarray} 
V = -{1 \over 2} \mu^2 \chi_{ij} \chi^{ij} \ + \ {1 \over 4}\lambda_1
(\chi_{ij} \chi^{ij})^2 \ 
+ \ {1 \over 4}\lambda_2 (\chi_{ij} \chi^{jk} \chi_{kl} \chi^{li}),
\end{eqnarray}
with $\chi_{ij} = -\chi_{ji} = (\chi^{ij})^*$. 
If $\lambda_2 > 0$ and $\mu^2 > 0$, then 
the vacuum expectation value (VEV) of $\chi$ is \cite{li}
\begin{eqnarray}
\langle \chi \rangle = c\left( \begin{array}{ccccc}
0 & & & &  \\
 & \left(\begin{array}{cc}
0 & 1 \\
-1 & 0 
\end{array}\right)  &  &  & \\
& & \left(\begin{array}{cc}
0 & 1 \\
-1 & 0
\end{array}\right)  & & \\
& & & \begin{array}{cc}
. &  \\
  & . 
\end{array} & \\
& & & & \left(\begin{array}{cc}
0 & 1 \\
-1 & 0
\end{array}\right) 
\end{array}
\right),
\end{eqnarray}
for $N$ odd with $c^2 = \mu^2/[\lambda_1 (N-1) + \lambda_2]$. 
For even $N$, the structure of the VEV is similar, except that there is no diagonal 
zero in the $(1,1)$ entry \cite{li}.  This VEV 
breaks $SU(N)_{\ell} \otimes U(1)_X$ to $Sp(N-1) \otimes U(1)_{Y}$ if $N$ is odd
and $SU(N)_{\ell} \otimes U(1)_X$ to $Sp(N)$ if $N$ is 
even, where, by definition, $Y \langle \chi \rangle = 0$. Evidently,
\begin{eqnarray}
Y = X + \left( {3y \over N}\right) T,
\label{Y}
\end{eqnarray} 
where $T$ is the diagonal $SU(N)$ generator 
\begin{eqnarray}
T = \left( \begin{array}{llllll}
-(N-1) & & & & &   \\
 & 1 & & & & \\
& & 1 & & & \\
& & & . & & \\
& & & & . & \\
& & & & & . 
\end{array}
\right)\ ,
\end{eqnarray}
and it is easy to see that $Y$ is standard hypercharge for the case $y=1/3$.

The correct low-energy phenomenology emerges if $N$ is odd (and $y = 1/3$), which means that the 
$N$ leptonic states split into the leptons, which do not gain any mass 
from $\langle \chi \rangle$, and an $Sp(N-1)$ $(N-1)$-plet of exotic heavy states 
which all gain mass from $\langle \chi \rangle$. In the special case of $N=3$, 
the three leptonic colours split up into the leptons and an 
$Sp(2) \simeq SU(2)'$ doublet of exotic fermionic states (called ``liptons'' 
in Ref.\cite{ql2}).

As mentioned above, with $N$ odd the model reduces to the standard model in the low energy limit,
provided $y = 1/3$. However, there is no theoretical explanation for the 
lightness of the neutrino masses or for why $y = 1/3$. We aim to address both of these 
issues with the physics associated with implementing the see-saw mechanism \cite{seesaw}.

We consider first the standard see-saw scenario with large Majorana masses for the $\nu_R$
induced through the VEV of a new Higgs multiplet $\Delta$:
\begin{eqnarray}
{\cal L} = \lambda \bar N_R \Delta (N_R)^c \ + \ H.c.
\end{eqnarray}
Clearly, 
\begin{eqnarray}
\Delta \sim \left({N(N+1) \over 2}, 1, 1, 
{-6y \over N} + 2\right),
\end{eqnarray}
and we require 
\begin{eqnarray}
\langle \Delta \rangle \propto
\left( \begin{array}{cccccc}
1& & & & & \\
 & 0 & & & & \\
 & & 0 & & & \\
& & & . & & \\
& & & & . & \\
& & & & & 0 
\end{array} \right).
\end{eqnarray} 
Note that if $y=1/3$ this VEV will break
\begin{equation}
SU(N)_{\ell} \otimes U(1)_X \to SU(N-1)_{\ell} \otimes U(1)_{Y},
\end{equation}
where $Y$ is standard weak hypercharge.  Since for even
$N-1$, $Sp(N-1)$ is a subgroup of $SU(N-1)$, the overall breaking
induced by $\langle \chi \rangle$ and $\langle \Delta \rangle$ is
to $Sp(N-1)\otimes U(1)_Y$.

As well as being responsible for the large Majorana $\nu_R$ mass, 
the $\Delta$ term can explain why $y=1/3$. To do this, the
interactions in the Higgs potential need to explicitly 
break one of the two anomaly-free 
Abelian symmetries, thereby removing all of the 
theoretical arbitrariness in the 
$U(1)_X$ charges.  (And of course the one left unbroken must be the correct one 
for generating the standard model at low energies.)

In particular, $SU(N)$ gauge invariance allows for terms which 
have one $\Delta$ and $N-1$ $\chi$'s (assuming $N$ is odd). 
For $N = 3$, the term is 
\begin{eqnarray}
V = \lambda \Delta_{ij} \epsilon^{ikl} \epsilon^{jmn} \chi_{kl} \chi_{mn}
\ + \ H.c.,
\end{eqnarray}
where we have made the $SU(3)$ indices explicit and $\epsilon$ is the 
completely antisymmetric $SU(3)$ invariant tensor.  
This term generalizes for $SU(N)$, as per
\begin{eqnarray}
V = \lambda \Delta \epsilon  \epsilon \chi^{N-1} \ + \ H.c.
\label{vt}
\end{eqnarray}
where, for simplicity, we have not made the $SU(N)$ indices explicit. 
This term explicitly breaks the second anomaly-free Abelian symmetry and fixes $y=1/3$.
Put another way, only for $y=1/3$ is the term in Eq.(\ref{vt}) gauge 
invariant.
However, this term is only renormalisable if it is of dimension four or 
less, which implies that $N = 3$ since $N$ is odd.
Thus, this type
of see-saw model can only be implemented for $N = 3$.

An upper limit on the scale of the new physics, $\langle \chi \rangle,
\langle \Delta \rangle$, arises if we are to avoid fine
tuning in the Higgs potential (gauge hierarchy problem).
In particular, Higgs mixing terms, $\lambda \phi^{\dagger} \phi Tr
\chi^{\dagger}\chi$ and  $\lambda' \phi^{\dagger}\phi Tr
\Delta^{\dagger}\Delta$ induce a shift, $\delta \mu^2$, of $\lambda Tr
\langle \chi \rangle^{\dagger} \langle \chi \rangle$ and $\lambda' Tr
\langle \Delta \rangle^{\dagger} \langle \Delta \rangle$,
in the $\mu^2$ coefficient of the quadratic $\phi^{\dagger}\phi$ term in
the Higgs potential. This suggests an upper limit of order 10 TeV for
$\langle \chi \rangle, \ \langle \Delta \rangle$, if we are to keep
$\mu^2$ at the electroweak scale naturally. Setting $\lambda, \lambda'
\to 0$ is not a viable option since these mixing terms are also induced
radiatively via gauge boson loops.

One problem with the standard see-saw mechanism is that the 
required light neutrino masses ($\stackrel{<}{\sim} eV$) suggest that
the $\nu_R$ 
scale should be much greater than 10 TeV.
This is theoretically problematic in view of the aforementioned gauge hierarchy 
problem.  To be
more precise about this, we should first be clear about the exact definition
of the usual see-saw model.  As well as the particle content and Lagrangian, the see-saw
model is defined by a specification of a certain parameter regime, namely
that the neutrino Dirac masses are of the same order as the charged fermion masses,
and that the right-handed Majorana masses are very much larger than that.
The same Lagrangian has, of course, other parameter regimes.  One of those other regimes
has the right-handed Majorana masses at the TeV scale or below, with correspondingly
much smaller neutrino Dirac masses.  That theory, which is {\it not} the see-saw model, 
does not have the same hierarchy
problem as does the see-saw model, because the highest scale is at worst not
far above the electroweak, and the very small neutrino Dirac masses are a technically
natural choice because the symmetry of the theory is increased as they are
taken to zero.  

To get around the hierarchy problem posed by the usual see-saw model, we consider an alternative
see-saw model \cite{wyler} that requires adding a family of gauge singlet 
fermions, $S_R \sim (1, 1, 1, 0)$,
to the spectrum. In this case there is no $\nu_R$ Majorana mass term, 
but instead there is a Dirac mass coupling $\nu_R$ and $(S_R)^c$. 
The neutral lepton mass matrix then has the form
\begin{eqnarray}
\left(\overline{\nu_L} \ \overline{(\nu_R)^c} \ \overline{(S_R)^c}\right) \left(
\begin{array}{ccc}
0 & m_D & 0 \\
m_D & 0 & m'_D \\
0 & m'_D & m'_m
\end{array}
\right)
\left( 
\begin{array}{c}
(\nu_L)^c \\
\nu_R \\
S_R 
\end{array}\right),
\end{eqnarray}
where $m'_m$ is a small Majorana mass term for the $S_R$ state. 
The neutrino mass in this model is then 
$m_{\nu} = {m_D^2 {m'}_m \over m_D^2 + {m'}_D^2 }$.
One elegant feature of this type 
of extended see-saw model is that the neutrino masses can be
naturally light, that is
without causing any gauge hierarchy problem. This is because $m'_m$
can be naturally small, as it arises from a bare mass term that when taken
to zero increases the symmetry of the theory.\footnote{In addition to the anomaly-free
$X$ charge, the Lagrangian also has the accidental anomalous (but
otherwise unbroken) global symmetries of
baryon number and, when $m'_m = 0$, (generalised) lepton number.  Taking
$m'_m \neq 0$ explicitly breaks the generalised lepton number.}
Taking $m'_m$
small is {\it not} conceptually similar to taking the neutrino Dirac masses to be small
in the usual see-saw case, because $m'_m$ has no connection with the
electroweak scale or any other spontaneous symmetry breaking scale.  In other
words, there is no {\it a priori} expectation that it should have any
particular value, and we are therefore free to make the parameter choice that
$m'_m$ is much less than a typical charged fermion mass.

To implement this extended see-saw scenario, we remove $\Delta$ and
in its stead introduce the scalar
$\Omega$ to induce the Dirac mass coupling $\nu_R$ and $(S_R)^c$:
\begin{eqnarray}
{\cal L} = \lambda \bar N_R (S_R)^c \Omega \  + \ H.c.
\end{eqnarray}
This means that $\Omega$ has the gauge quantum numbers
\begin{eqnarray}
\Omega \sim \left(N, 1, 1, {-3y \over N} + 1\right),
\end{eqnarray}
and we require the VEV
\begin{eqnarray}
\langle \Omega \rangle \propto \left( \begin{array}{c}
1\\
0 \\
0 \\
. \\
. \\
0 \end{array}\right).
\end{eqnarray}
If $y=1/3$, this VEV performs the symmetry breaking
\begin{equation}
SU(N)_{\ell} \otimes U(1)_X \to SU(N-1)_{\ell} \otimes U(1)_{Y},
\end{equation}
where $Y$ is standard weak hypercharge $Y$.  Since for even
$N-1$, $Sp(N-1)$ is a subgroup of $SU(N-1)$, the overall breaking
induced by $\langle \chi \rangle$ and $\langle \Omega \rangle$ is
to $Sp(N-1) \otimes U(1)_{Y}$.

As in the previous case (the minimal see-saw model without $S_R$), 
we use terms in the Higgs potential to explicitly break the second anomaly-free Abelian symmetry 
and so to fix $y=1/3$. In this case, the Higgs potential term is
\begin{eqnarray}
V = \lambda \Omega \epsilon \chi^{(N-1)/2}\ + \ H.c.
\end{eqnarray}
where $N$ is odd and, for simplicity, we have not made the $SU(N)$ 
indicies explicit.\footnote{For example, 
in the case of $N=3$, $V = \lambda \Omega_i \epsilon^{ijk} \chi_{jk}$.}
Again, this term is gauge invariant if and only if $y=1/3$.
Renormalisability requires that 
the scalar operator in $V$ must be of dimension four or less, which implies
that $(N-1)/2 \le 3$, and thus there are three renormalisable models
corresponding to $N = 3, 5, 7$.  (For the case $N = 3$, the field $\Omega$ need
not be independent, as its quantum numbers are identical to $\chi^*$\footnote{The
original $N=3$ model \cite{fv95} employing the mechanism of Ref.\cite{wyler} suffers
from a proton decay problem, due to the necessary presence of a discrete quark-lepton
symmetry partner to the Higgs field $\chi$, 
as discussed recently in Ref.\cite{nuquart}.
The models proposed here do not have this problem, because there 
is no discrete
symmetry and hence no need to introduce 
a quark partner for $\chi$, even for the
$N=3$ case.}).
It is theoretically most natural 
if $\langle \chi \rangle, \langle \Omega \rangle \stackrel{<}{\sim}$ 10 TeV 
in order to avoid the gauge hierarchy problem.
This means that the new physics associated 
with the $SU(N)\otimes U(1)_X/Sp(N-1)\otimes U(1)_Y$
coset space gauge bosons and liptons can be accessible to future colliders such as
the LHC. Flavour-changing neutral-current processes such as 
$\mu \to e\bar e e, \ \mu \to e \gamma$ are also 
induced and can be close to the experimental limit for a TeV symmetry 
breaking scale \cite{ql2}. 

Perhaps the most characteristic new physics predicted by these types of 
models is the phenomenology associated with the $N-1$ exotic leptonic
coloured degrees of freedom -- liptonic physics.  The liptons have 
electric charges $\pm 1/2$, but are confined into two-particle bound states by 
the $Sp(N-1)$ gauge interactions. These bound states have electric charges
$0, \pm 1$.  Their phenomenology has been studied in detail in Ref.\cite{ql2} 
for the specical case of $N=3$. Here we recall and generalise some
of the interesting features of this phenomenology. 

We use the following notation for the liptons. The $N$ leptonic colours
of $SU(N)_{\ell}$ shall be expressed as:
\begin{eqnarray}
F_L = \left( \begin{array}{c}
f_L \\
F_{1L} \\
F_{2L} \\
.\\
.\\
F_{(N-1)L}
\end{array}
\right), \
E_R = \left( \begin{array}{c}
e_R \\
E_{1R} \\
E_{2R}\\
. \\
. \\
E_{(N-1)R} \end{array}
\right), \
N_R = \left( \begin{array}{c}
\nu_R \\
V_{1R} \\
V_{2R} \\
. \\
. \\
V_{(N-1)R} \end{array}
\right)
\end{eqnarray}
where $f_L, e_R, \nu_R$ are the familiar leptons and right-handed
neutrino, and $F_{iL}, E_{iR}, V_{iR}$ $(i=1,2,...,N-1)$ are the exotic fermions 
we call liptons.  The $SU(2)_L$ degrees of freedom are denoted by
\begin{eqnarray}
f_L = \left( \begin{array}{c}
\nu_L \\ e_L 
\end{array}\right), \ 
F_{iL} = \left(\begin{array}{c}
V_{iL}\\ E_{iL} 
\end{array} \right).
\end{eqnarray}

Let us start by focusing on the bound states made up of the lightest
lipton, $L_1$.  The $Sp(N-1)$ gauge interactions have an $SU(2)$ global flavour
symmetry group, which we denote by $SU(2)_F$.  
The lipton, $L_1$, and its anti-particle, $L_1^c$ 
transform as a doublet under this $SU(2)_F$ flavour group:
\begin{eqnarray}
L= \left( \begin{array}{c}
L_1 \\
L_1^c 
\end{array}
\right).
\end{eqnarray}
Of course, $L_1$ and $L_1^c$ have the same mass, so the $SU(2)_F$
symmetry is not broken by any mass difference. The flavour group $SU(2)_F$ is 
broken only by the electromagnetic and other electroweak interactions.
The $Sp(N-1)$ gauge interactions will confine the liptons into two-particle 
bound states.  If $\Lambda_{Sp(N-1)} \ll M_{L_1}$ [where $\Lambda_{Sp(N-1)}$
is the $Sp(N-1)$ confinement scale], then these liptonic bound 
states will be non-relativistic. 

The flavour structure of the hadrons follows from the $SU(2)_F$ decomposition
\begin{eqnarray}
2 \otimes 2 = 1_A \oplus 3_S,
\end{eqnarray}
where the subscripts $S, A$ denote the symmetry property under
interchange of the liptons.
The wave function of the liptons in the bound state can be expressed
as
\begin{eqnarray}
\psi = \psi_{colour} \otimes \psi_{space} \otimes \psi_{flavor} \otimes
\psi_{spin}
\end{eqnarray}
Here ``colour'' refers to the $Sp(N-1)$ quantum numbers, and is anti-symmetric.
The ground state has zero orbital angular momentum ($\Rightarrow$ $\psi_{space}$ is symmetric), 
and thus the Pauli principle implies that the flavour triplet will have spin-1 and the flavour 
singlet will have spin-0.  We denote these states by
\begin{eqnarray}
\rho^+ &=& L_1 L_1, \ \rho^0 = {L_1 L_1^c + L_1^c L_1 \over \sqrt{2}}, \
\rho^- = L_1^c L_1^c\nonumber \\
\xi^0 &=& {L_1 L_1^c - L_1^c L_1 \over \sqrt{2}}.
\end{eqnarray}
These particles can be produced from, for example, virtual $W$, $Z$ and $\gamma$
decays in future colliders such as the LHC. Furthermore, the production cross 
section will have an $Sp(N-1)$ `colour' factor, enabling the number of colours to
be experimentally determined.
Note that all of these bound states can decay into the standard particles,
with distinctive decay modes such as $\xi^0 \to \gamma \gamma, \rho^- \to e \bar \nu, 
\rho^0 \to e\bar e$ \cite{ql2}.  Reference \cite{ql2} contains an extensive discussion
of phenomenological bounds on the $N=3$ model, concluding that the scale of the
new physics can be as low as a few TeV. This result is not expected to be much
altered by more recent data, and we also expect the $N=5,7$ models to also be
phenomenologically acceptable for leptonic colour breaking scales of a few TeV and above.

In conclusion, we have explored the possibility that there is an
$SU(N)_{\ell}$ colour gauge group for leptons generalising the familiar
$SU(3)_q$ colour gauge group of
the quarks. We have shown that the see-saw mechanism for small neutrino
masses, along with some other theoretical constraints (electric charge quantisation)
suggest that the models with $N = 3, 5, 7$ are theoretically the most interesting. 
A striking feature of these types of models is the physics 
associated with the extra leptonic degrees of freedom: the liptons. These 
particles can potentially be discovered at future colliders such 
as the LHC, making these models testable in the near future.

\vskip 1.5cm
\noindent
{\bf Acknowledgements:}
This work was supported by the Australian Research Council.

\end{document}